\documentclass{iaus}
\usepackage{graphicx}
\usepackage{iaus-patch}

\title[Star formation in the Magellanic Clouds] 
{The star formation process in the \\ Magellanic Clouds}

\author[Oliveira]   
{J.M. Oliveira}

\affiliation{Lennard-Jones Laboratories, School of Physical \& Geographical
Sciences, Keele University, Staffordshire ST5 5BG, UK \\email: {\tt
joana@astro.keele.ac.uk}}

\pubyear{2008}
\volume{256}  
\jname{The Magellanic System: Stars, Gas, and Galaxies}
\editors{Jacco Th. van Loon \& Joana M. Oliveira, eds.}

\begin{document}

\maketitle

\begin{abstract}
The Magellanic Clouds offer unique opportunities to study star formation both
on the global scales of an interacting system of gas-rich galaxies, as well as
on the scales of individual star-forming clouds. The interstellar media of the 
Small and Large Magellanic Clouds and their connecting bridge, span a range in 
(low) metallicities and gas density. This allows us to study star formation
near the critical density and gain an understanding of how tidal dwarfs might
form; the low metallicity of the SMC in particular is typical of
galaxies during the early phases of their assembly, and studies of star
formation in the SMC provide a stepping stone to understand star
formation at high redshift where these processes can not be directly observed.
In this review, I introduce the different environments encountered in the
Magellanic System and compare these with the Schmidt-Kennicutt law and the
predicted efficiencies of various chemo-physical processes. I then concentrate 
on three aspects that are of particular importance: the chemistry of the 
embedded stages of star formation, the Initial Mass Function, and feedback 
effects from massive stars and its ability to trigger further star formation.
\keywords{astrochemistry, stars: formation, stars: luminosity function, mass 
function, ISM: clouds, H\,{\sc ii} regions, ISM: molecules, galaxies: evolution,
Magellanic Clouds, galaxies: stellar content}
\end{abstract}

\firstsection 
\section{Introduction}

The Magellanic Clouds (MCs) are our closest gas-rich galaxy neighbours. New 
instrumental advances mean we can now study their resolved stellar populations 
in great detail. At the same time, in the MCs we can study these populations 
from outside the galaxies themselves, allowing us a unique view of these 
populations and how they relate to the gas and dust distributions and the 
galaxies' structure.

The formation of stars in the early Universe took place in a metal-poor 
environment, however most of what is currently known about the star formation
process is derived from observations in the Milky Way. A great advantage of the 
MCs is that their Interstellar Mediums (ISM) are characterised by metallicity 
significantly lower than that of the Milky Way. Thus the MCs are ideal templates
to test whether metallicity significantly influences star formation and thus are
unique probes for the environmental conditions more typical of the 
high-redshift Universe.

Only since relatively recently are we able to study the details of the star
formation in the MCs. Firstly, I describe the star formation environment in the
Magellanic Clouds. I will then concentrate on three particular 
facets of the star formation process. I will start by discussing the effects of a
lower metal content on the chemistry in molecular clouds. I will then discuss
the stellar Initial Mass Function (IMF), as well as massive star feedback and 
triggered star formation at low metallicity.

\section{The star formation environment in the MCs}

In the Milky Way star formation is clearly dominated by the spiral arms
(\cite[see review by Elmegreen 2009]{elmegreen09}). In galaxies with weak or no
spiral arms, star formation seems to occur throughout their disks, probably
resulting from local gravitational instabilities. This is the case of the MCs,
that exhibit no clear spiral arms but are still observed to be forming stars at 
present\footnote{Although the LMC can be considered a one-armed spiral (see 
Wilcots, these proceedings), star formation in the LMC is not limited to this 
particular structure.}. On the other hand, the MCs are interacting gas-rich 
galaxies that also interact with the Milky Way. Thus tidal and/or hydrodynamical 
effects might have an important role in stimulating and regulating star 
formation in these galaxies. The properties and location of the young stellar 
populations identified in the SMC tail (the part of the Magellanic Bridge 
closest to the SMC) suggest that these stars formed within a body of gas that
had already been pulled out of the SMC, probably by tidal forces 
(\cite[Harris 2007]{harris07}). Whether these tides were also responsible for 
triggering the star formation is not clear. It has been proposed that the 
formation of the giant complex 30\,Doradus could be associated with the last 
Magellanic collision, about 0.2\,Gyr ago (\cite[Bekki \& Chiba 2007]{bekki07}),
which may have induced a gaseous spiral arm distortion lasting more than an 
orbital period. The alternative explanation of ram-pressure induced star 
formation as a result of the LMC ploughing through the hot Galactic halo 
(\cite[de Boer et al. 1998]{deboer98}) has recently gained support again,  
explaining also some features of the Magellanic Stream 
(\cite[Nidever, Majewski \& Burton 2008]{nidever08}).

\begin{figure}
\begin{center}
 \includegraphics[width=13cm]{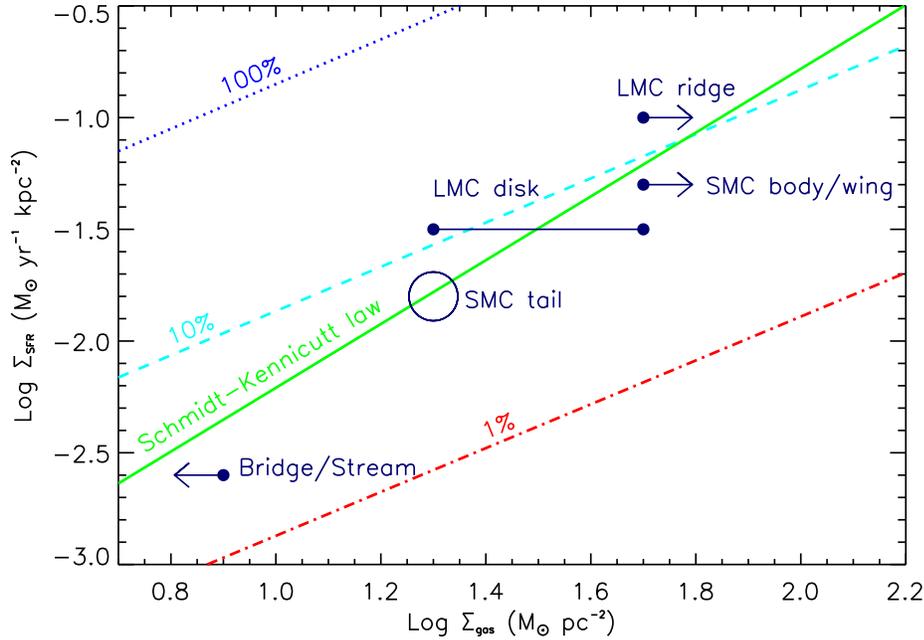} 
 \caption{Schmidt-Kennicutt law (\cite[Kennicutt 1998]{kennicutt98}) with gas
 densities for different regions of the Magellanic System indicated (estimated
 from \cite[Nidever et al. 2008]{nidever08}, see 
 Table\,\ref{table1}). Different global star formation efficiencies are also 
 indicated (\cite[Kennicutt 1998]{kennicutt98}).}
\label{fig0}
\end{center}
\end{figure}

The star formation efficiency is generally observed to be correlated with the
local gas density. This Schmidt law (\cite[Schmidt 1959]{schmidt59}) --- and its
generalization to include starbursts by \cite[Kennicutt (1998)]{kennicutt98} ---
is usually expressed in terms of the projected star formation rate and gas 
column density. It is observed that the efficiency of star formation is lower 
when the gas density is higher, and there appears to be a lower threshold to the
gas density that can support star formation 
(\cite[Kennicutt 1998]{kennicutt98}). Typical gas densities for distinct 
components in the Magellanic System may be estimated from the H\,{\sc i} data 
presented in \cite[Nidever et al. (2008, see their Fig.\, 9)]{nidever08}, where 
the H\,{\sc i} column density was transformed into total gas density assuming 
that the atomic and molecular gas fractions are equal (see Fig.\ 4 in 
\cite[Kennicutt 1998]{kennicutt98}), a common average distance of 60\,kpc, and 
ignoring projection effects. These are listed in Table\,\ref{table1} and 
compared with the Schmidt-Kennicutt law in Fig.\,\ref{fig0}. Note that the 
molecular fraction of gas in the Bridge/Stream is likely to be smaller.

\begin{table}[h]
\begin{center}
\caption{Relevant properties of different regions in the Magellanic System.
Typical gas densities are estimated from 
\cite[Nidever et al. (2008)]{nidever08}.}
\label{table1}
\vspace{2mm}
\begin{tabular}{|c|c|r|c|c|}
\hline
region & $Z/{\rm Z_{\odot}}$& \multicolumn{1}{c|}{N(H\,{\sc i})}& $\Sigma$(gas) & star formation\\
& &atoms cm$^{-2}$ & M$_{\odot}$ pc$^{-2}$ &\\
\hline
LMC ridge & 0.4& $>$ 3 $\times$ 10$^{21}$ & $>$ 50 & yes/soon\\
LMC disk & 0.4 & 1$-$3 $\times$ 10$^{21}$ & \llap{$\sim$}\,20$-$50 & yes, scattered\\
SMC body/wing & 0.2 & $>$ 3 $\times$ 10$^{21}$ & $>$ 50 & yes, throughout\\
SMC tail & 0.1 & 1($-$2) $\times$ 10$^{21}$ & $\sim$ 20 & yes, in pockets\\
Bridge/Stream & ? & $<$ 5 $\times$ 10$^{20}$ & $\ll$ 8 & no\\
\hline
\end{tabular}
\end{center}
\end{table}

The MCs offer a range of gas densities, with the SMC tail being a unique example
of a star forming entity close to the critical density, while high-density 
pockets such as 30\,Dor and much of the generally very gas-rich SMC main body 
support vigorous star formation. The Magellanic Bridge and Stream are apparently
too tenuous as they show no signs of star formation. The transition between 
quiescent and star-forming ISM within the Magellanic System seems to occur for a
gas column density at which 
\cite[Krumholz, McKee \& Tumlinson (2008)]{krumholz08} predict that molecular 
clouds become self-shielded. It has now become possible to examine the 
Schmidt-Kennicutt law on pc scales within individual star-forming clouds in the 
MCs (see Indebetouw, these proceedings), with the aim of identifying deviations 
from the global law resulting from differences in the physical processes that 
work on small scales within galaxies.

The Magellanic Clouds have a low metallicity ISM (0.1\,$-$\,0.4\,Z$_{\odot}$, 
Table\,\ref{table1}). This implies that there is in general less dust in 
the MCs than in the Milky Way, i.e. the gas-to-dust ratio is higher. As we will 
see in the next section, this might have important implications for the star 
formation process. The efficiency of some chemo-physical processes are expected 
to depend on metallicity. Fig.\ 2 shows typical timescales for several of these
processes that are expected to play an important role in the star formation 
process and their dependence on metallicity (from 
\cite[Banerji et al. 2008]{banerji08}). This diagram shows that the relative 
hierarchy of these processes (i.e. which ones might be more dominant due to 
shorter timescales) seems to change even in the metallicity range covered by the
Magellanic System and the Milky Way. This suggests that metallicity might have 
an effect on the star formation process, even without having to go back to the 
conditions prevalent when the very first generations of stars formed.

\begin{figure}[t]
\begin{center}
 \includegraphics[width=13cm]{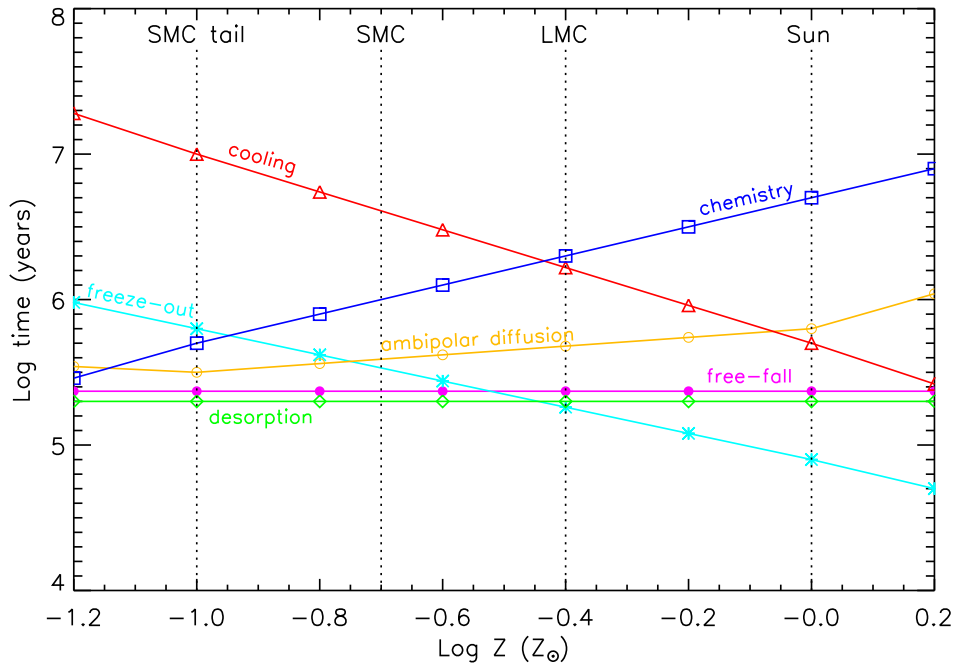} 
 \caption{Timescales relevant for the star formation process as a function of
 metallicity (adapted from \cite[Banerji et al. 2008]{banerji08}). Processes
 considered are: free-fall collapse (filled circles), freeze-out (crosses; 
 formation of ice coatings on dust grains), ambipolar diffusion (open circles),
 desorption (diamonds; process by which molecules formed in ice mantles are 
 returned to the gas phase), cooling (triangles) and ion-molecule chemistry 
 (squares; creation of coolant molecules from atomic gas). Some of these 
 processes and their timescales depend on metallicity. In the metallicity range 
 covered by the MCs and the Galaxy, the relative importance of these processes 
 seems to change.}
\label{fig1}
\end{center}
\end{figure}

\section{Star formation and molecular cloud chemistry}

Why is chemistry important for the early stages of the star formation process? 
First of all chemistry is essential in determining the cooling in a contracting 
molecular clump. During the onset of gravitational collapse of a dense cloud, 
sufficiently dense cores can only develop if the heat produced during the 
contraction can be effectively dissipated. This cooling occurs mainly via 
fine-structure lines of C or O, for instance via the C\,{\sc ii} emission line 
at 158\,$\mu$m and also via rotational transitions of abundant molecules like 
water. At lower metallicity there are obviously fewer C and O atoms and it is
expected that this might somehow affect the cooling efficiency
(Fig.\,\ref{fig1}). 

Magnetic fields may also play an important role in supporting dense molecular
clouds and the cores within them. Magnetic field lines are coupled directly to 
ions and, via ion-neutral collisions, are also coupled to the neutral material.
If this latter coupling is weak, then the neutral material decouples from the
magnetic field lines and can then respond more freely to any gravitational
perturbation (ambipolar diffusion). With fewer metallic atoms available in low 
metallicity environments, the ionization state of the gas as a whole could be 
lower and collapse may proceed closer to free fall. On the other hand, less dust
shielding and harsher radiation field (see below) might cause metallic atoms to 
be ionized to a higher degree. Furthermore, it is not clear how important 
ambipolar diffusion really is 
(\cite[Crutcher, Hakobian \& Troland 2008]{crutcher08}).

In cold molecular clouds, gas-phase, ice and dust chemistries are strongly 
inter-linked. Dust grains play an important role by shielding the cold gas in 
the molecular clouds against ambient UV radiation. Dust grains also provide 
surfaces onto which chemical reactions occur that would otherwise not be 
possible. For instance surface chemistry is important in the formation of both
H$_{2}$ and O$_{2}$ molecules. Ice mantles form on the surfaces of dust grains,
leading to depletion of molecules from the gas phase.

In the MCs the gas-to-dust ratio is higher. If there is less dust then we can 
expect that the shielding effect of dust opacity is weaker, compounded by the 
fact that the ambient UV radiation in MCs is harsher than that in our Galaxy
(\cite[Welty et al. 2006]{welty06}). Chemistry in general might be slower if 
there is less grain surface available. Extinction curves are different in 
the diffuse ISM in the MCs (\cite[Gordon et al. 2003]{gordon03}), possibly due 
to a different grain size distribution. If dust grains were predominantly 
smaller in molecular clouds in the MCs, they would provide more surface per unit
mass, possibly counteracting the effect of a lower total dust mass. In the 
denser environment of star forming clouds, grains grow and the rate at which 
this occurs may be different at low metallicity. 

Dust composition is possibly also different. It has been suggested that there 
might be more carbon-rich dust produced by evolved stars in the SMC
(\cite[Zijlstra et al. 2006]{zijlstra06}). It is not clear what effect this 
would have in dust composition as carbon-rich dust is more easily destroyed (cf.
\cite[van Loon et al. 2008]{vanloon08}). Nevertheless, if dust composition is 
different in the MCs, one could expect a different opacity and consequently a 
different thermal balance. In summary, if the properties of dust grains are 
intrinsically different in the MCs this could affect chemistry in general and, 
via cooling, the physics of the star formation process.

\subsection{Ice chemistry}

In cold star forming clouds, ice mantles form on the surface of dust grains. 
Abundant molecules like water, CO$_{2}$ (and to some extent CO) are largely 
locked into these ice mantles (\cite[Bergin et al. 1995]{bergin95}). The most 
abundant ice species are water (typically $10^{-5}-10^{-4}$ with respect to 
H$_{2}$), followed by CO$_{2}$ and CO, with combined abundance 10$-$30\% with 
respect to water ice (\cite[van Dishoeck 2004]{vandishoek04}). Ice processing 
also enriches the gas phase: molecules like methanol, formaldehyde and formic 
acid are possibly formed via UV and cosmic ray processing of ice mantles before
being evaporated into the gas phase. Processing also causes segregation of the 
different ice species and crystallinisation within the ice mantles. Therefore 
understanding ice chemistry is a powerful route to understand cloud chemistry 
in general. 

\begin{figure}
\begin{center}
 \includegraphics[width=9.cm,angle=-90]{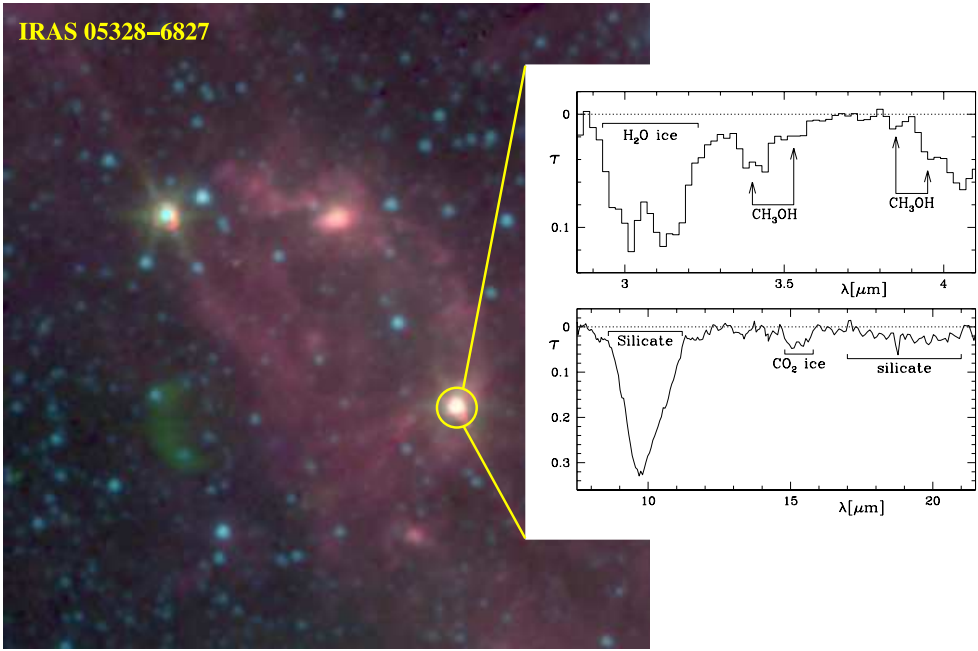} 
 \caption{IRAS\,05328$-$6827 in the LMC, the first extra-galactic embedded YSO 
spectroscopically identified by \cite{vanloon05}. The image is a 3-colour IRAC 
image; the spectra clearly show absorption features due to water and CO$_{2}$ 
ices as well as silicate dust.}
\label{fig2}
\end{center}
\end{figure}

\begin{figure}
\begin{center}
 \includegraphics[width=13.2cm]{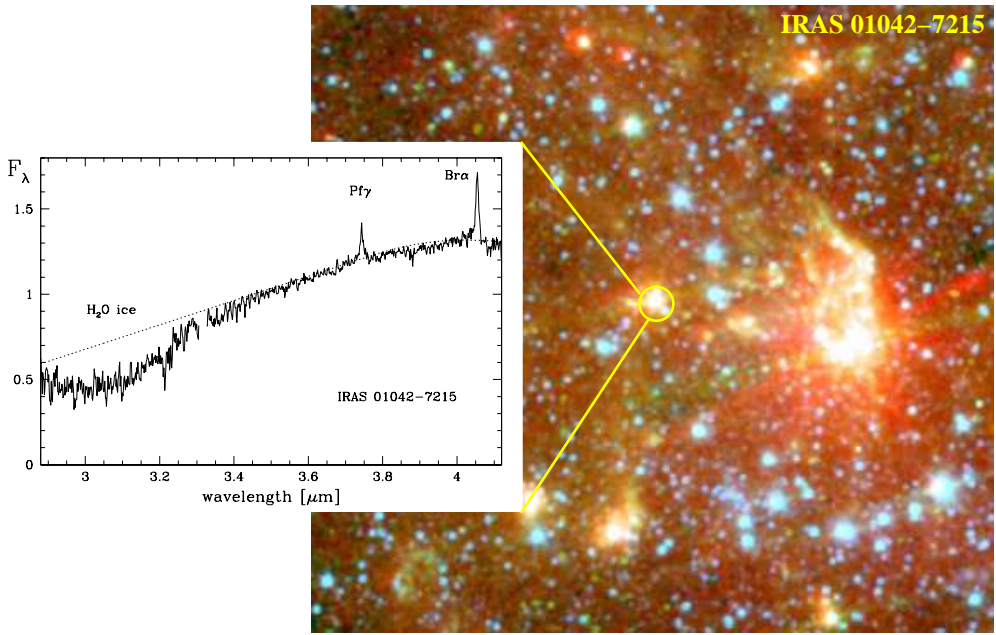} 
 \caption{IRAS\,01042$-$7215 in the SMC, a YSO identified in the SMC
(\cite[van Loon et al. 2008]{vanloon08}). The image is a 3-colour IRAC image; 
the spectra clearly show absorption features due to water ice, as well as 
emission lines, indicating a more evolved object than IRAS\,05328$-$6827
(Fig.\,\ref{fig2}).}
\label{fig3}
\end{center}
\end{figure}

Ices have been detected in the envelopes of heavily embedded young stellar 
objects (YSOs) both in the Milky Way and the MCs. The first evidence of 
ices in the envelope of a massive YSO in any extra-galactic environment was a 
serendipitous discovery by \cite{vanloon05}. The Spitzer IRS and ISAAC/VLT 
spectrum of IRAS\,05328$-$6827 in the LMC shows clear absorption signatures of 
water ice at 3.1\,$\mu$m, methanol ice at 3.5 and 3.9\,$\mu$m and CO$_{2}$ ice 
at 15.2\,$\mu$m (Fig.\,\ref{fig2}). The spectrum also shows a typical silicate 
dust absorption feature at 10\,$\mu$m. Recently \cite{shimonishi08} identified 
a few more embedded YSOs with ice signatures in the LMC. \cite{vanloon05} 
suggests that ice processing observed in IRAS\,05328$-$6827 might be an effect 
of metallicity, as UV radiation is able to deeper penetrate the less dusty
envelopes of metal-poor YSOs. This is a tentative result and a more extensive 
sample, preferably in the SMC, would be needed to constrain any definitive 
metallicity trend. Three embedded YSOs have been identified in the SMC 
(\cite[van Loon et al. 2008]{vanloon08}); the ISAAC/VLT spectrum of one of these
sources, IRAS\,01042$-$7215, shows clear water ice absorption and hydrogen 
emission lines, indicating a slightly more evolved object than the LMC example 
(Fig.\,\ref{fig3}).

What are the possible effects of low metal content on the chemistry in the
envelopes of  massive YSOs? Lower carbon and oxygen abundances in the 
metal-poor ISM of the MCs lead to lower gas-phase CO abundances 
(\cite[Leroy et al. 2007]{leroy07}). One could therefore also expect a 
lower CO and CO$_2$ content of the ice mantles that coat the dust grains. On 
the other hand a lower dust fraction leads to reduced shielding from an already
harder interstellar radiation field. A high degree of ice processing could thus 
be expected as well as a higher degree of crystallinity and ice species 
segregation. While the first effect could be expected to depress the abundances 
of all species when compared to Galactic embedded YSOs, the second effect would 
enhance more complex ice species mainly at the expense of water ice. It is not 
clear at present which, if any, of these effects dominates. By comparing YSO 
samples in the SMC, LMC and the Milky Way we can probe metallicities from 
0.1\,$-$\,1\,Z$_{\odot}$, allowing us to assess the effect of metallicity on 
ice chemistry. 

Firstly, we will have to construct reliable samples of YSOs in the MCs, selected
in a more systematic way, to be compared with Galactic samples. The Spitzer 
Space Telescope imaging surveys of the MCs --- SAGE 
(\cite[Meixner et al. 2006]{meixner06}), S$^3$MC 
(\cite[Bolatto et al. 2007]{bolatto07}) and SMC-SAGE
(\cite[Gordon et al., in preparation]{gordon08}) --- have identified large 
numbers of YSO candidates (\cite[Whitney et al. 2008]{whitney08}). Follow-up 
spectroscopic surveys (including SAGE-SPEC and SMC-IRS) target a sizeable YSO 
sample, that will allow the study of ice features in the mid-IR range. AKARI is 
also being used to investigate ice features on many such objects
(\cite[Shimonishi et al. 2008]{shimonishi08}) in the near-IR, as are groundbased
investigations (\cite[Oliveira et al., in preparation]{oliveira09}). 

Observations of Galactic YSOs show that the bands of the main ice species are 
better fitted by laboratory profiles that include admixtures of other ices
(\cite[Pontoppidan et al. 2008]{pontoppidan08}). Furthermore, from the previous 
paragraph it is clear that relative ice abundances, not individual ice species, 
hold the key to isolate any environmental chemistry effect. Another important 
point to take into account is that in the MCs we can only expect to study the 
most massive embedded YSOs, thus one needs to carefully consider the objects' 
luminosity before comparing the samples. The analysis of the spectral energy 
distributions of such objects allows to constrain not only the objects' 
luminosity and evolutionary stage but also dust properties that together with 
ice chemistry will build a coherent picture of the early star formation process.

One of the major advances since the previous MC Symposium has been the
availability of facilities like Spitzer which now allow us to study in detail 
the properties of embedded YSOs outside our Galaxy. We can thus sample star 
formation environments that are significantly different from those encountered 
in the Milky Way, with the real possibility of understanding how metallicity 
influences the early stages of star formation. At the same time, we can now also
study global properties of the resulting young stellar populations and 
investigate for instance how the Initial Mass Function (IMF) reflects their 
parent environment. 
 
\section{From cloud collapse to the stellar IMF}

The stellar IMF is determined by a range of inter-linked physical processes:
cooling, turbulence, fragmentation, feedback, rotation, magnetic fields etc. The
IMF is a rather simplistic snapshot of a very complicated process, and we still 
struggle to fully understand it and explain its main properties.

At the extreme low metallicities of the early Universe, the IMF is believed to
have been top heavy, i.e\, star formation events gave rise only to extremely 
massive stars, due mainly to the inability of primordial gas to efficiently cool
at low temperatures (\cite[Abel, Bryan \& Norman 2002]{abel02}). As these 
populations die and new ones emerge, the Universe gets progressively enriched 
with copious amount of metals. Once metallicities of the order of 10$^{-3}$ to 
10$^{-5}$\,Z$_{\odot}$ are reached, enough metals are present to allow for more 
efficient cooling, through fine-structure and molecular transitions, as well as 
continuum emission from dust produced in supernova explosions. This more 
efficient cooling allows the gas that will form the next generation of stars to 
reach lower temperatures, and therefore smaller clumps can be created via 
fragmentation (\cite[e.g. Smith \& Sigurdsson 2007]{smith07}). That is to say 
lower-mass stars can form and the peak of the IMF 
shifts to lower masses. Thus, once the metal content in the Universe reaches 
this metallicity threshold stellar populations are characterised by essentially 
a Salpeter-like IMF with a turn-over or characteristic mass at lower masses, as
observed in the solar neighbourhood (\cite[Kroupa 2001]{kroupa01};
\cite[Chabrier 2005]{chabrier05}).

The present day local IMF seems to be universal, but significant variations 
are observed mainly in more extreme environments, like regions that formed at 
high redshift (\cite[Elmegreen 2008]{elmegree08a}). The IMF can be defined as
$\xi (\log M) \propto M^\Gamma$, where $\Gamma$ is the IMF slope.
Observationally, the IMF is found to have a slope of $\Gamma \sim -1.35$ 
(commonly referred to as the ``Salpeter slope'', 
\cite[Salpeter 1955]{salpeter55}) down to below a solar mass; it then 
flattens out somewhat (the so-called IMF plateau) and reaches a peak at about 
0.3\,M$_{\odot}$, falling sharply into the brown dwarf regime. Theoretical 
considerations actually struggle to explain the observed IMF constancy in a 
wide variety of environments (\cite[Kroupa 2007]{kroupa07}). However, this 
refers mostly to the slope of the IMF at higher masses or to integrated stellar 
populations; only very recently are we able to probe directly the low-mass IMF 
in environments that may be distinct to those prevalent in the Milky Way.

In what way could we expect metallicity to influence the lower mass IMF? Based 
on numerical simulations it has been proposed that the characteristic mass 
scales with the thermal Jeans mass at the onset of collapse 
(\cite[Bate \& Bonnell 2005]{bate05}). One could then intuitively expect the 
IMF properties to vary with the environmental conditions, via for instance the 
dependence of the Jeans mass on the temperature and metallicity in the molecular
core. However, \cite{elmegreen08b} show that, if grain-gas coupling is taken 
into account, the thermal Jeans mass should depend only weakly on environmental
factors like density, temperature, metallicity and radiation field. In 
particular, the dependence of the Jeans mass with metallicity could be only as 
$Z^{-0.3}$. It is also possible that the characteristic mass is not linked to 
the Jeans mass, depending instead on core sub-fragmentation and protostellar 
feedback (\cite[Elmegreen 2008]{elmegreen08a}).

What have we learned so far from constraining the IMF in the MCs? One of the 
first regions investigated to characterise the high-mass IMF outside our Galaxy 
was R\,136 in the LMC. In the early days, it was thought that R\,136 was a 
single super-massive star 
(\cite[Cassinelli, Mathis \& Savage 1981]{cassinelli81}), suggesting an IMF with
different properties to those observed in the Galaxy. Modern instrumentation 
however quickly resolved R\,136 into a massive star cluster 
(\cite[Weigelt \& Baier 1985]{weigelt05}). In fact, the most massive stars 
observed in the MCs have masses of the order of $\sim$\,150\,M$_{\odot}$
(\cite[Massey \& Hunter 1998]{massey98}), consistent with Galactic massive 
objects. We should point out that these are observed masses. As discussed in 
\cite{elmegreen08a}, very massive stars loose a substantial amount of mass 
extremely quickly. Thus this observed maximum mass is not necessarily the 
maximum mass as far as the star formation process is concerned.

The IMF of several associations in the MCs is found to have a Salpeter slope
down to about 1\,$-$\,2\,M$_{\odot}$ (\cite[Sirianni et al. 2000]{sirianni00};
\cite[Selman \& Melnick 2005]{selman05}; 
\cite[Kumar, Sagar \& Melnick 2008]{kumar08} to name but a few). To study the 
lower-mass IMF, we need to investigate populations younger than about 10\,Myr, 
as these are less affected by dynamic evolution effects. Young solar-mass stars 
were firstly identified using near-IR images or H$\alpha$ emission, for instance
in 30\,Doradus (\cite[Brandl et al. 1996]{brandl96}) and near SN\,1987\,A 
(\cite[Panagia et al. 2000]{panagia00}). More recently a number of Hubble Space 
Telescope (HST) surveys have allowed the detection of large numbers of 
pre-main-sequence (PMS) objects in H\,{\sc ii} regions in the Clouds, down to 
below half a solar mass (e.g. \cite[Nota et al. 2006]{nota06}; 
\cite[Gouliermis et al. 2006, 2007]{gouliermis06,gouliermis07a}). 

Two young star forming regions in the SMC have been recently under very 
close scrutiny (see also contributions by Gouliermis and Sabbi in this volume). 
NGC\,602 in the wing of the SMC shows that star formation can be triggered in 
low-density environments, probably by compression and turbulence associated with
H\,{\sc i} shell interactions (\cite[Nigra et al. 2008]{nigra08}). The massive 
cluster in the center has many PMS stars associated with it 
(\cite[Gouliermis, Quanz \& Henning 2007]{gouliermis07b}) and embedded ongoing 
star formation has been detected towards the rim of the H\,{\sc ii} region 
(\cite[Carlsson et al. 2007]{carlsson07}). NGC\,346 on the other hand is the 
largest OB association in the SMC and it has a more complex gas and dust 
morphology than NGC\,602. A rich PMS has been discovered in this region 
(\cite[Nota et al. 2006]{nota06}) and star formation is still ongoing in 
its denser parts (\cite[Simon et al. 2007]{simon07}). It has been proposed that 
several star formation episodes have taken place in NGC\,346
(\cite[Sabbi et al. 2007]{sabbi07}; 
\cite[Hennekemper et al. 2008]{hennekemper08}). For both these young regions the
IMF is found to be Salpeter-like down to about a solar mass 
(\cite[Schmalzl et al. 2008]{schmalzl08}; \cite[Sabbi et al. 2008]{sabbi08}); 
below that mass incompleteness becomes a serious issue. 

Even with HST imaging we are unable to fully sample the IMF plateau and 
determine the characteristic mass of young stellar populations in the MCs. But
even when new instrumentation allows us to do so, we might struggle to find 
conclusive answers. If the characteristic or peak mass depends on metallicity as
proposed by \cite{elmegreen08b}, in the MCs we could expect it to be about 
0.4$-$0.6\,M$_{\odot}$, where in the Galaxy it typically is 0.3\,M$_{\odot}$.
Furthermore, even in the Milky Way we observe variations in the peak mass that 
seem to be significant (\cite[Luhman et al. 2007]{luhman07}; 
\cite[Oliveira et al. 2008]{oliveira08}) however we are as yet unable to explain
these differences. Another possible complication is that the peak mass also 
depends (albeit weakly) on the density and radiation field, and in the MCs we 
do sample lower densities and harsher radiation fields. Therefore a metallicity 
effect on the IMF properties might prove difficult to isolate.

\section{Feedback and triggering}

Another factor we have to take into account is that star formation has a cyclic
component to it: feedback from massive stars can quench/prolong a star formation
episode and maybe even trigger new such events. If feedback effects are
different at low metallicity, can this affect the outcome of a star formation
event?

Infant massive stars start to shape their environments very early in their
evolution, by ionizing the gas and heating the dust 
(\cite[Oey \& Clarke 2007]{oey07}). Stellar UV radiation and winds create 
expanding blisters of ionized gas. At the interfaces between these H\,{\sc ii} 
regions and the cold molecular cloud, the heating powers shocks that compress 
the molecular material. Supernova explosions also disrupt severely the 
neighbouring ISM. In the first instance, massive stars violently destroy their 
parent molecular clouds. However, observational evidence indicates that at the
rims of giant H\,{\sc ii} regions star formation is ongoing, NGC\,6611 and the
Eagle Nebula in the Milky Way being one such example (\cite[Oliveira 2008]
{oliveira08b}). This seems to suggest that under the right conditions, radiative
and/or mechanical feedback of massive stars can trigger further star formation 
in the remnant molecular gas (see review by Chu in this volume).

\begin{figure}[t]
\begin{center}
\includegraphics[width=6.65cm,bbllx=-93bp,bblly=82bp,bburx=570bp,bbury=713bp,clip=]{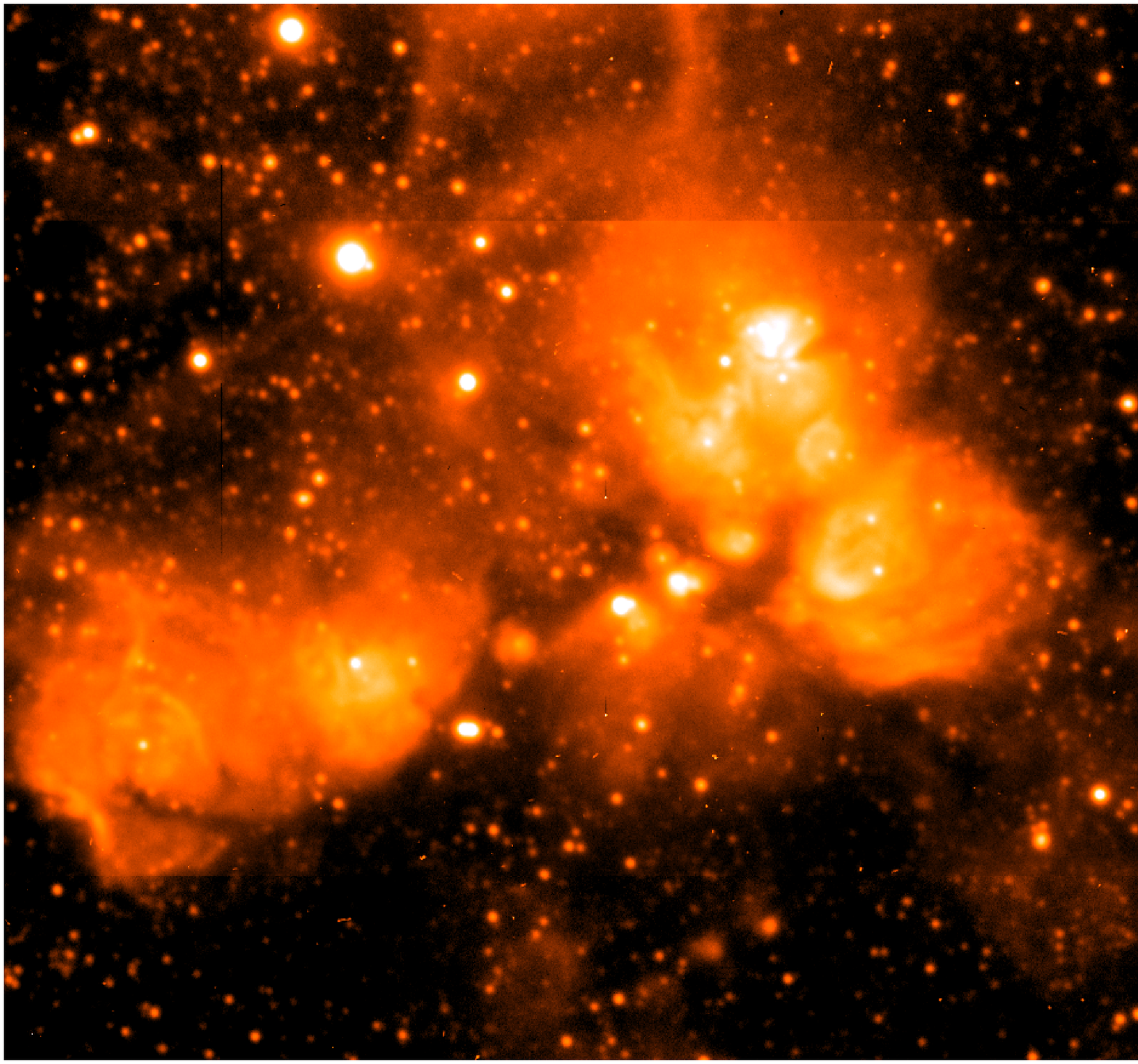}
\hspace*{-1.1mm}\includegraphics[width=6.65cm,bbllx=-93bp,bblly=82bp,bburx=570bp,bbury=713bp,clip=]{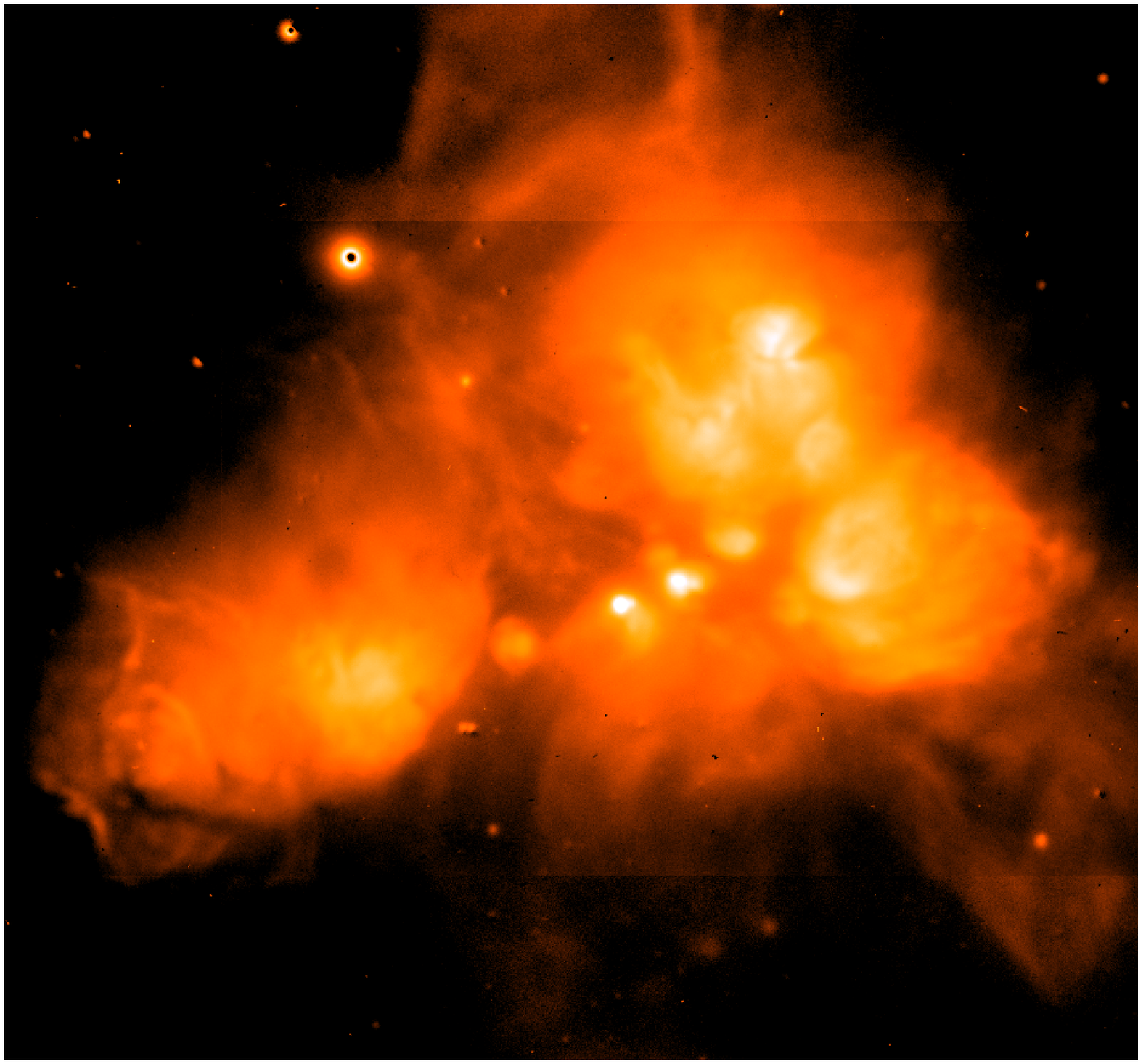}
\vspace*{-1mm}\includegraphics[width=6.65cm,bbllx=72bp,bblly=162bp,bburx=540bp,bbury=600bp,clip=]{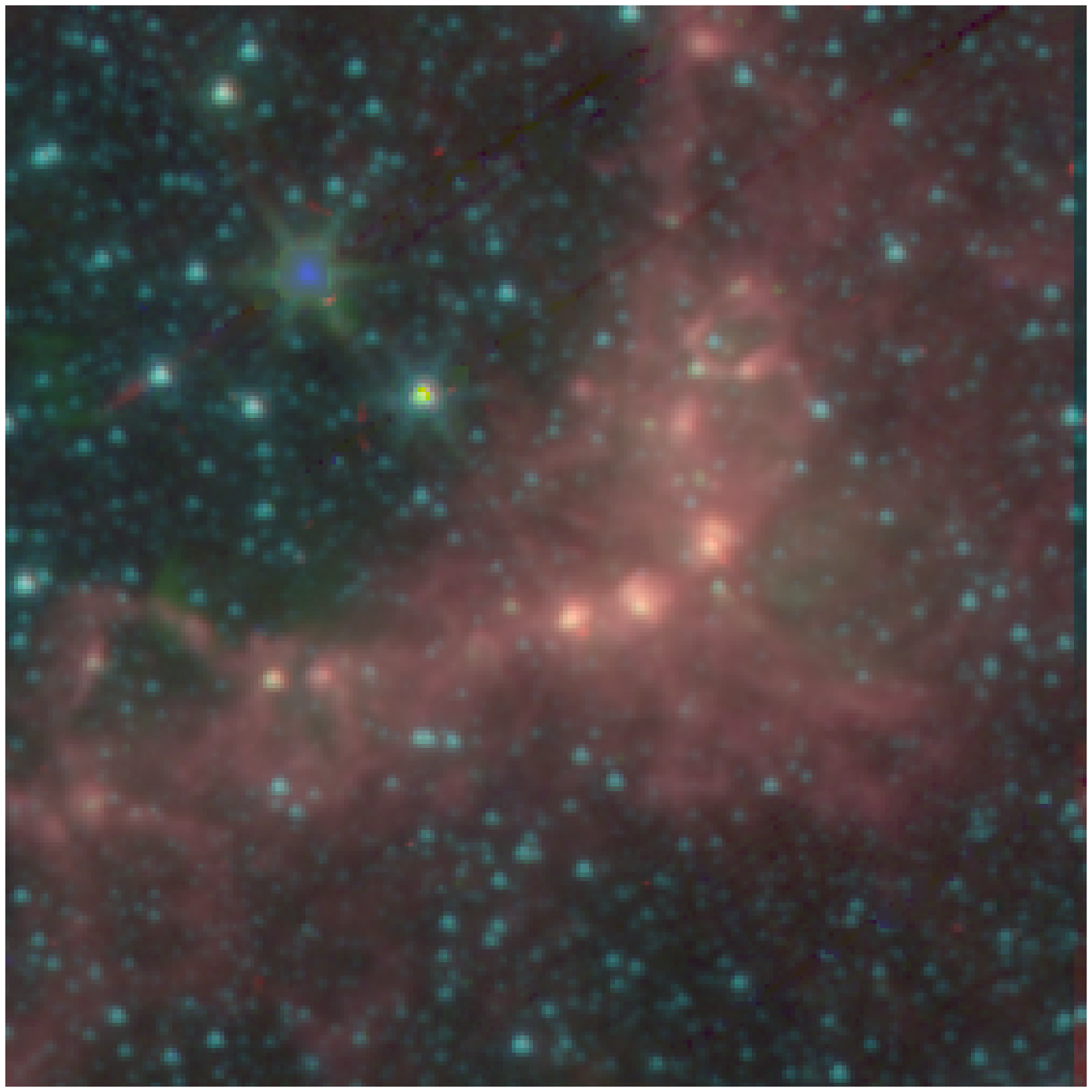}
\hspace*{-1.1mm}\includegraphics[width=6.65cm,bbllx=72bp,bblly=162bp,bburx=540bp,bbury=600bp,clip=]{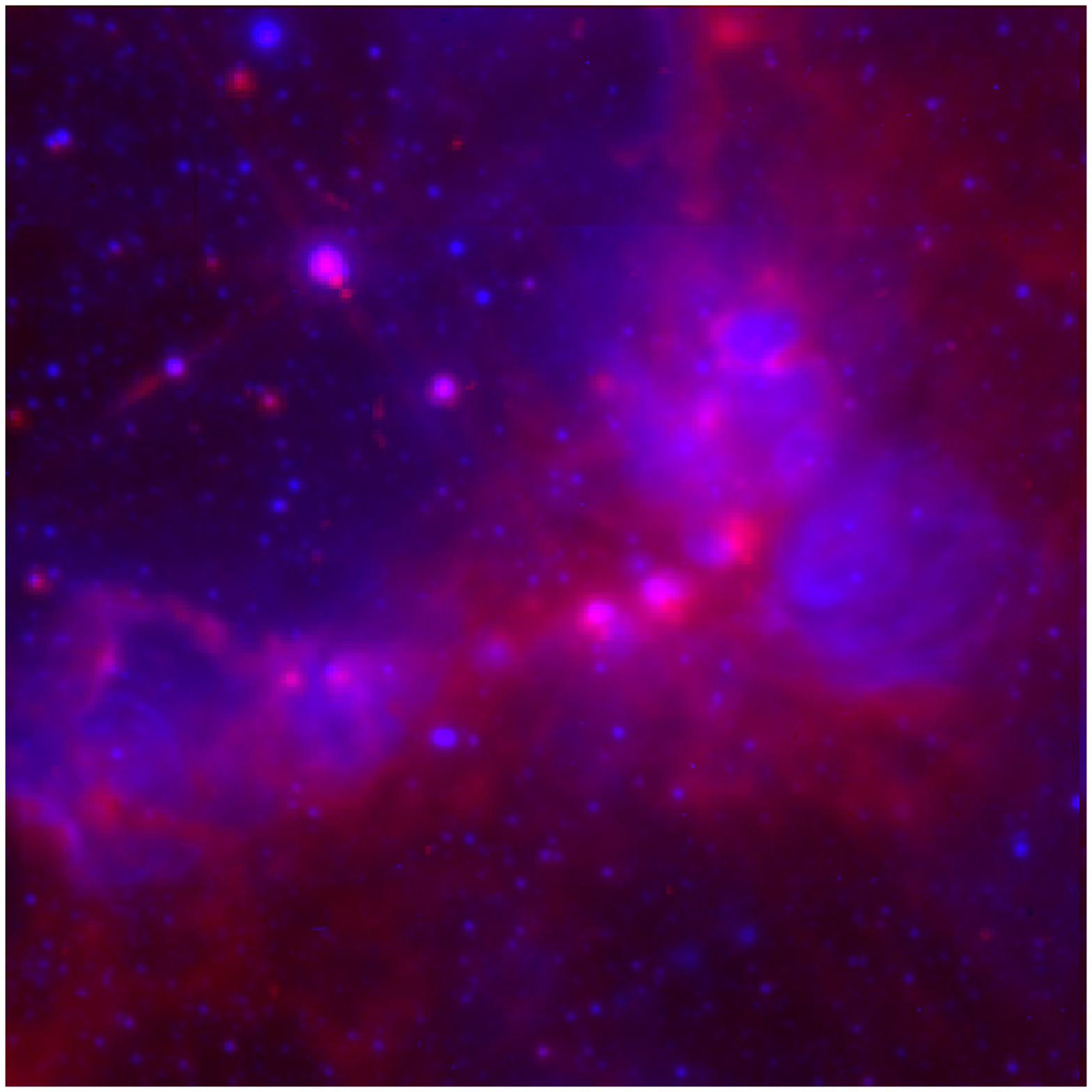}
\caption{Multi-wavelength images of N\,113 in the LMC 
(\cite[Oliveira et al. 2006]{oliveira06}). The images at the top are the 
H$\alpha$+continuum (left) and H$\alpha$ emission (right), showing the 
complex structure of the hot gas. The image on the bottom left shows an 
IRAC/Spitzer 3-colour composite, showing the location of the dense dust lane
in the center of the region. The image on the bottom right shows a composite of 
H$\alpha$+continuum with IRAC/Spitzer 8\,$\mu$m dust emission. It shows to 
dramatic effect how the bubbles of ionized gas created by the massive stars are 
compressing the dense lanes of dust. Star formation is ongoing in the compressed
dust, as shown by the presence of maser emission 
(\cite[Oliveira et al. 2006]{oliveira06}).}
\label{fig4}
\end{center}
\end{figure}

The observed properties of massive stars in the Magellanic Clouds show some 
important differences when compared to their galactic counterparts (see review 
by Evans in this volume). Massive stellar winds are slower and less dense in the
low metallicity environment of the MCs (\cite[Mokiem et al. 2007]{mokiem07}).
For the same stellar mass, massive stars in the Clouds are also hotter than 
their Galactic counterparts (\cite[Massey et al. 2005]{massey05}), which could
compensate slightly the metallicity effect on the wind as hotter stars drive
faster outflows. The lower dust-to-gas ratio in the MCs implies that dust 
shielding is less efficient and radiative feedback from massive stars may 
thus be more effective at lower metallicity. All this seems to suggest that 
massive star feedback might operate differently at low metallicity. But do we 
know how this affects the triggering of star formation?

N\,113 is a small H\,{\sc ii} region in the LMC. Multiple young stellar 
populations are associated with this region 
(\cite[Bica, Claria \& Dottori 1992]{bica92}) and star formation is ongoing 
within radio continuum sources in the central part of the nebula 
(\cite[Wong et al. 2006]{wong06}), as pinpointed by the detection of water and 
OH maser emission (\cite[Lazendic et al. 2002]{lazendic02}; 
\cite[Brooks \& Whiteoak 1997]{brooks97}). Fig.\,\ref{fig4} shows in detail the
interplay between the ionized gas and the molecular material in N\,113.
\cite{oliveira06} use H$\alpha$ emission and continuum images together with
Spitzer/IRAC images to show that the ionized gas bubbles created by 
the massive stars in the region are compressing the dense lanes of molecular
material where star formation is ongoing. Further persuasive evidence for 
triggered star formation in the MCs is found at the rims and interfaces of LMC 
supergiant shells where ongoing star formation is occurring (see contribution by 
Chu, this volume). This paints the picture that star formation triggered by 
massive stars and their interaction with the ISM also occurs in the Clouds.

One problem with analysing the processes that can trigger star formation is to 
establish a cause-effect relationship: it is very difficult to distinguish 
between sequential star formation that occurs without the need for extra
triggers and star formation that occurs only due to the direct action of the 
massive stars. One of the few examples where triggering may have been proved is 
the superbubble N\,51D in the LMC. By measuring the thermal pressure in a dust
globule and comparing it to ambient conditions, \cite{chu05} find that star 
formation in the globule may have been induced recently by the thermal pressure 
in the superbubble interior. 

As we have seen in Section\,2, in the MCs there are other mechanisms that could 
trigger star formation: ram pressure and/or tidal effects resulting from 
interactions within the Magellanic System and with the Milky Way. Even though 
we know that massive stars have different observed properties at low 
metallicity, it is not clear whether this significantly influences star
formation triggering.

\section{Final remarks} 
 
I reviewed what we know about the star formation environment in the 
Magellanic Clouds, in particular concerning gas density and metallicity. 
I described how particular components of the star formation process, namely,
the chemistry of the early embedded stages, stellar IMF and stellar feedback and
triggering, might operate differently in environments with low metal abundance 
as in the Magellanic Clouds. The Magellanic System offers an ideal laboratory to 
study both the details and global properties of star formation at low 
metallicity. This kind of studies has only recently become possible; with so 
much new data and exciting facilities coming on, we have very much to learn!

\end{document}